\documentclass[submission,copyright,creativecommons]{eptcs}

\providecommand{\event}{GandALF 2024}
\usepackage{scalerel}
\usepackage{paralist}
\usepackage{amsmath}
\usepackage{cancel}
\usepackage{wrapfig}
\usepackage{changepage}
\usepackage{tcolorbox}
\usepackage[linesnumbered,ruled,vlined]{algorithm2e}

\usepackage{subcaption}
\usepackage{makeidx}
\usepackage[linguistics]{forest}
\usepackage{float}
\usepackage{amsthm}
\usepackage[T1]{fontenc}
\usepackage{paralist}
\usepackage{graphicx}
\usepackage{amsfonts}
\usepackage{comment}
\usepackage{xparse}

\newcommand{\cv}{\check{v}}
\newcommand{\hv}{\hat{v}}

\newcommand{\bbN}{\mathbb{N}}

\newcommand{\bbR}{\mathbb{R}}

\newcommand{\cB}{\mathcal{B}}
\newcommand{\cC}{\mathcal{C}}
\newcommand{\cD}{\mathcal{D}}

\newcommand{\cP}{\mathcal{P}}

\newcommand{\cT}{\mathcal{T}}

\newcommand\before[1][\le]{\ensuremath{#1}}

\DeclareDocumentCommand\DeclareTemporalRelation{m O{\le} m}{%
  \DeclareDocumentCommand#1{}{\before[#2]^{\mathsf{#3}}}%
}

\DeclareTemporalRelation\sbefores{s,s}
\DeclareTemporalRelation\ebeforee{e,e}
\DeclareTemporalRelation\sbeforee{s,e}
\DeclareTemporalRelation\ebefores{e,s}
\DeclareTemporalRelation\sbefore{s}
\DeclareTemporalRelation\ebefore{e}
\DeclareTemporalRelation\safter[\ge]{s}
\DeclareTemporalRelation\eafter[\ge]{e}
\DeclareTemporalRelation\startsat[=]{s}
\DeclareTemporalRelation\endsat[=]{e}

\newcommand{\cI}{\mathcal{I}}

\usepackage{iftex}

\newtheorem{theorem}{Theorem}
\newtheorem{definition}{Definition}
\newtheorem{problem}{Problem}

\newcommand{\spin}{\mathsf{SPIN}}

\usepackage{tikz}
\usetikzlibrary{decorations.pathreplacing, calc, fit}

\usepackage{fontawesome}
\usepackage{stix}
\usepackage{xspace}

\usepackage{amsmath}

\newcommand{\oT}{\overline{T}}

\newcommand{\CPI}{\textsf{BPMN+CPI}}

\newcommand{\incoming}{\scalebox{0.8}{$incoming$}}
\newcommand{\outgoing}{\scalebox{0.8}{$outgoing$}}

\newcommand{\mnce}{\scalebox{0.8}{$\textit{MNCE}$}}
\newcommand{\places}{\scalebox{0.8}{$\textit{Places}$}}
\newcommand{\outplaces}{\scalebox{0.8}{$\textit{OutPlaces}$}}

\ifpdf
  \usepackage{underscore}         
  \usepackage[T1]{fontenc}        
\else
  \usepackage{breakurl}           
\fi

\title{Reactive Synthesis for Expected Impacts}
\author{Emanuele Chini
\institute{Department of Computer, Control\\ and Management Engineering,\\ University ``La Sapienza'',\\ Rome, Italy.
\email{emanuele.chini@uniroma1.it}}
\institute{Department of Computer Science,\\ University of Verona, Verona (Italy)
\email{emanuele.chini@univr.it}
}
\and Pietro Sala
\institute{Department of Computer Science,\\
University of Verona, Verona (Italy)
\email{pietro.sala@univr.it}
}
\and
Andrea Simonetti
\institute{Department of Computer Science,\\
University of Verona, Verona (Italy)
\email{andrea.simonetti@studenti.univr.it}
}
\and Omid Zare
\institute{Department of Computer Science,\\
University of Verona, Verona (Italy)
\email{omid.zare@univr.it}
}
}
\def\titlerunning{Reactive Synthesis for Expected Impacts}
\def\authorrunning{E. Chini, A. Simonetti, P. Sala, and O. Zare}

\usepackage{lineno}

\hypersetup{
  bookmarksnumbered,
  pdftitle    = {\titlerunning},
  pdfauthor   = {\authorrunning},
  pdfsubject  = {EPTCS},               
  pdfkeywords = {Business Process Management, Expected Impacts, Reactive Synthesis} 
}

\begin{document}
\maketitle

\begin{abstract}

As business processes become increasingly complex, 
effectively modeling decision points, their likelihood, 
and resource consumption is crucial for optimizing operations. 
To address this challenge, this paper introduces a formal 
extension of the Business Process Model and Notation (BPMN) 
that incorporates choices, probabilities, and impacts, 
referred to as \CPI. This extension is motivated 
by the growing emphasis on precise control within 
business process management, where carefully 
selecting decision pathways in repeated instances 
is crucial for conforming to certain standards of multiple resource consumption and environmental impacts. 
In this context we deal with the problem of synthesizing a 
strategy (if any) that guarantees that the expected impacts
on repeated execution of the input process 
are below a given threshold.  We show that this problem belongs to  
PSPACE complexity class; moreover we provide an effective procedure 
for computing a strategy (if present).

\end{abstract}

\section{Introduction}\label{sec:intro}
BPMN (Business Process Model and Notation) has emerged as a pivotal formalism in the realm of process management, offering a standardized method for detailing business processes in various sectors, including healthcare and industry. Its graphical notation facilitates the clear and precise representation of process flows, enabling stakeholders to comprehend, analyze, and improve business operations. 
In the healthcare sector, BPMN plays a critical role in implementing patient care guidelines~\cite{pufahl2022bpmn}. 
Similarly, in the industrial domain, it aids in the efficient management of manufacturing and supply chain processes, ensuring timely delivery of products and services~\cite{fernandes2021role}.  In these domains, increasing attention has arisen in the past decade on the topic of Business Processes Management, where the choice of traces on the control side is paramount. These applications demand measurement and employ, as a means for selection, notions such as \textit{cost-awareness}\cite{magnani2007computing}, \textit{energy-awareness}\cite{cappiello2010first}, and \textit{resource-awareness}~\cite{de2012data}, which naturally induce scenarios where multiple measurements must be controlled.

In this paper, we proceed under the implicit assumption that all costs, energies, and resources utilized are positive and exhibit additive characteristics. This implies that our process instances solely deplete resources to fulfill their objectives without the capability to generate resources. As we will demonstrate, this restriction contributes to favorable computational properties.

Moreover, we use the probabilistic split, referred to as \emph{nature}, which signifies a decision based on a probability distribution beyond the worker’s control. For instance, in healthcare, a \emph{nature} is the chance of developing gastritis when taking Brufen 600 with a probability of $1\%$.
Similarly, in industrial applications, machinery wear and tear may influence the production process, requiring maintenance stops during production. 

Finally, time consumption for tasks is considered, as they will be equipped with specific durations.

Our approach here is twofold. First, we aim to introduce a formal BPMN extension that addresses execution in the presence of all the previously mentioned components, namely, BPMN plus Choices/ Probability/Impacts (\CPI). Next, we seek to provide a dynamic control mechanism, i.e., a strategy, for BPMN execution. This is to ensure, where possible, that the expected impacts remain below a set of user-defined thresholds. 
To elegantly juggle all these concepts within a single framework, we enrich the standard Petri Net semantics for BPMN to capture impacts, durations, and probabilities. We call this model of computation the Simultaneous Probabilistic Impactful Network ($\spin$). We then define a graph representing all possible executions of $\spin$. This graph is combined with a natural modification of classical reachability games to derive the desired strategy, if any. The primary aim of this study is to determine, for a process formalized in \CPI, whether a controller exists that can accurately execute each step of the process while ensuring that the expected value of each resource, across repeated process instances, remains within predefined thresholds.

Upon establishing the computational model for \CPI, we tackle the challenge of synthesizing a strategy for a specified process in \CPI, given a set of expected value thresholds. This is achieved through the following steps:

\begin{compactenum}
\item \emph{Semantics by Petri Nets}. After defining how to translate a \CPI\ into a $\spin$, we define the semantics of both of them by giving the semantics of $\spin$ alone as an extension of classical Petri net semantics. This includes introducing time durations for places, probabilistic transitions, and the possibility (under certain conditions) of executing a set of enabled transitions simultaneously instead of one at a time;

\item \emph{Computation Graph}. All possible computations for the given $\spin$ are represented as a graph. In this graph, each node represents a path of executions, and any edge between two computations indicates that the source computation can be extended to the target computation by firing one or more enabled transitions in the source computation;

\item \emph{Classical Reachability Game Graph Transformation}~\cite{thomas1995synthesis}. By transforming the computation graph into a classical reachability game graph, where spoiler nodes (typically denoted by $\square$) represent choices made by nature, we assess the existence of a “good” set of final states that can “attract” the initial state. If such a set exists, we can infer the existence of our strategy.
\end{compactenum}

The paper is organized as follows. In Section~\ref{sec:related}, we present and describe related work and the state-of-the-art algorithms for finding strategies in computational models that can encode \CPI\ through suitable translation. In Section~\ref{sec:cpi}, we illustrate a practical example of a BPMN process in an industrial setting, followed by a formal definition of the \CPI\ model, detailing the components of choices, probabilities, and impacts. Since we restrict ourselves to acyclic graphs, at the end of this section, we briefly discuss a simple way of dealing with loops within the proposed framework. In Section~\ref{sec:complexity}, we provide the complexity bounds for the strategy synthesis problem for \CPI.
While Section~\ref{sec:complexity} deals with the decision problem of establishing whether a strategy exists or not, Section~\ref{sec:strategies} focuses on effectively synthesizing a strategy given a \CPI\ process and a bound for expected impacts. Finally, 
Section~\ref{sec:conclusion} highlights our main findings, their theoretical and practical impacts, and future research avenues.

\section{Related work}\label{sec:related}
\begin{table}[t]
\centering
\scalebox{0.9}{
\begin{tabular}{llll}
\multicolumn{1}{c}{\textbf{Methods}} & \multicolumn{1}{c}{\textbf{Costs}}                                                   
& \multicolumn{1}{c}{\textbf{Durations}}                                          & \multicolumn{1}{c}{\textbf{Strategy}}                                                                                                                                  \\ \hline
UPPAAL-Stratego                       & \begin{tabular}[c]{@{}l@{}}multiple,\\ not considered\\ for strategy\end{tabular}                              & \begin{tabular}[c]{@{}l@{}}explicitly defined, \\ time is continuous\end{tabular} & \begin{tabular}[c]{@{}l@{}} $\bullet $non-deterministic \\ $\bullet$ state explosion due to subset construction\end{tabular}                                                             \\ \hline
PRISM                                & \begin{tabular}[c]{@{}l@{}}multiple \\ negative allowed\end{tabular}                                                      & \begin{tabular}[c]{@{}l@{}}implicit via \\ multiple states\end{tabular}         & \begin{tabular}[c]{@{}l@{}} $\bullet$ $\epsilon$-approximated  strategy \\  $\bullet$ increases exponentially  
w.r.t $1/\epsilon$
\end{tabular}                                       \\ \hline
MPG-MDP                              & \begin{tabular}[c]{@{}l@{}}multiple \\ negative allowed\end{tabular}           & \begin{tabular}[c]{@{}l@{}}implicit via \\ multiple states\end{tabular}         & \begin{tabular}[c]{@{}l@{}} $\bullet$ infinite plays\\ $\bullet$ \CPI\
would need difficult encoding\\ $\bullet$ game averages values on a per-step basis\end{tabular}                                             \\ \hline
\textbf{Our method}                                 & \begin{tabular}[c]{@{}l@{}}multiple, \\only positive\end{tabular}                                  & \begin{tabular}[c]{@{}l@{}}explicitly defined \end{tabular}      & \begin{tabular}[c]{@{}l@{}} $\bullet$ deterministic \\ $\bullet$ exact strategy by integrating rewards\\ and probabilities\\ 
$\bullet$ game averages values on a per-instance basis\end{tabular} \\ \hline
\end{tabular}
}
\caption{A summary of the features of the tool introduced in this study and the problems addressed by UPPAAL-Stratego, PRISM, and MPG-MDP, respectively.}
\label{tab:comparison}
\vspace{-0.5cm}
\end{table}

The most commonly accepted semantics for BPMN processes, used for both formal tasks like monitoring, verification, and querying, and application-driven tasks like process discovery and execution forecasting, is the Petri Net semantics. In this approach, a BPMN process is mapped into a Petri Net \cite{dijkman2008semantics}. This mapping retains several beneficial properties, including the crucial feature that the resulting net is 1-bounded~\cite{SymbAlgoSyntesisPN}, meaning that from an initial state with one token, all configurations will have at most one token per place. Under this 1-boundedness assumption, the Petri Net reduces to an exponentially succinct representation of a finite automaton (FA) \cite{reachparamonecounterautomata}, where all labelings can be represented as sets of places holding one token, making the number of states finite. If the language of this automaton is defined by its transitions, the resulting FA is deterministic (DFA). Thus, many formal problems, such as querying, emptiness checking, strategy synthesis (reachability games), and Linear Temporal Logic (LTL) model checking, can be equivalently viewed in BPMN, 1-bounded Petri Nets, or succinct DFAs, as transformations between these representations can be performed in LOGSPACE.

Incorporating resources into BPMN processes is well-explored in process optimization literature. In \cite{magnani2007computing}, an extension of the classical BPMN notation is proposed to evaluate the overall cost of process diagrams, comparing costs associated with tasks as single values or intervals to find the most cost-effective way to perform the intended job. Our contribution specifically focuses on the positive impacts of such integration, further allowing the specification of impacts as arrays of cost values to express monetary costs and other resources or requirements.
In \cite{combi2019modular}, Combi et al. outlined a method for enforcing distinctive temporal behaviors by introducing temporal patterns (e.g., minimum and/or maximum durations) linked to tasks. They proposed creating reusable, duration-aware process models using existing BPMN elements, capturing duration constraints at various abstraction levels, and checking for duration constraint violations at runtime.
Duran et al. \cite{duran2018stochastic} introduced a rewriting logic executable specification of BPMN extended with time and probabilities, allowing stochastic expressions to specify task durations and flow delays.
Herbert~et~al.~\cite{herbert2013precise} formalize an extension of the BPMN language incorporating probabilistic nondeterministic branching. Additionally, they present an algorithm for translating such models into MDPs expressed in the syntax of the PRISM model checker~\cite{KNP11}.
This facilitates precise quantitative analysis of business processes. 
We have adopted a similar extension of BPMN to introduce non-deterministic behaviour (for nature nodes), which is frequently observed in real-world application scenarios. Probabilities are linked to gateway branching behaviors, enabling discrete-event simulation and automatic stochastic verification of various properties. Our work will consider task durations by imposing stringent time constraints, ensuring that each task extends over a time interval precisely equal to its duration, which possibly affects which choice is enabled first in a given execution. Additionally, incorporating probabilities into BPMN situates our research within the specialized domain of Markov Decision Processes (MDPs)~\cite{filar2012competitive}, significantly enhancing the applicability of BPMN in decision-making under uncertainty.

Moving beyond BPMN, our approach primarily involves devising a strategy within an MDP enhanced with vectors of positive impacts. The objective is to ensure that the strategy’s expected value does not exceed a specific threshold. The realm of strategy synthesis for MDPs has been extensively explored, leading to notable breakthroughs like the PRISM model checker~\cite{KNP11}. PRISM has emerged as a key instrument, evolving over time to incorporate sophisticated features for strategizing within MDP contexts.
Another notable development is UPPAAL-Stratego~\cite{UPPAAL}, an extension of the well-regarded UPPAAL-TIGA~\cite{behrmann2007uppaal}, which solves the strategy synthesis problem for games played on timed automata incorporating both costs and probabilities. From a theoretical perspective, albeit less focused on specific tools, our issue shares similarities with Mean Payoff Games (MPG)\cite{ZWICK1996343} as applied to MDP (MPG-MDP)\cite{chatterjee2011energy}.
The differences and similarities between our proposed method and the current state of the art are concisely summarized in Table~\ref{tab:comparison}. While we focus on a system with probabilities, we are aware of other formalisms that allow impact vectors with negative contributions, such as infinite energy games~\cite{DBLP:conf/csl/AbdullaAHMKT14}.

\section{\CPI: Processes with Choices, Probabilities, and Impacts}\label{sec:cpi}

In this section, we begin by informally illustrating the concept of \CPI\ through an intuitive example of a metal manufacturing process together with an initial, intuitive understanding of the expected impacts induced by a strategy in Section~\ref{subsec:example}. These concepts are then formalized in Section~\ref{subsec:problem}, where we also state the core problem of this work: finding an optimal strategy that minimizes the overall impact. Finally, in Section~\ref{subsec:loops}, we discuss the advantages and drawbacks of reducing diagrams with loops to acyclic ones from the perspective of strategy synthesis. 

\subsection{Motivating Example}\label{subsec:example}

\begin{figure}
    \centering
    \includegraphics[scale= 0.4]{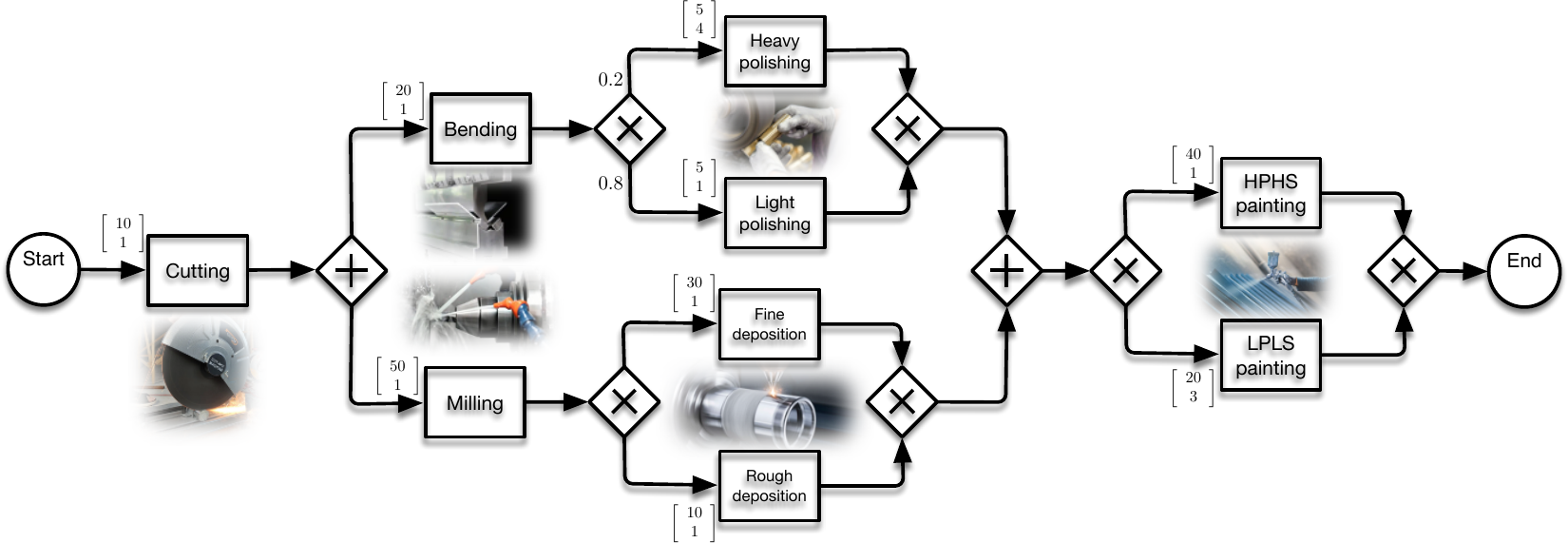}
    \caption{An example of \CPI\ diagram for an industrial process.}
    \label{fig:motivatingexample}
\end{figure}

The \CPI\ diagram of Figure~\ref{fig:motivatingexample} depicts a metal manufacturing process that involves cutting, milling, bending, polishing, depositioning, and painting a metal piece. It consists of a single-entry-single-exit (SESE) diagram, with a choice, a nature, and an impact for each task, which is defined as a numbers vector.  
The bracketed numbers next to each activity represent impact vectors
$ \scriptsize \begin{bmatrix}
a \\
b \end{bmatrix}$ where $a$ = cost of the task and $b$ = hours/men required to complete the task. For instance, cutting the metal piece costs 10 units (e.g., currency, resource, etc.), and requires 1 unit of time or manpower (e.g., 1 hour or 1 worker).
In Figure~\ref{fig:motivatingexample}, the nature's probability of each chosen path is indicated with the numbers next to decision points. For example, there’s a high probability (0.8) of the process moving from bending to light polishing and a low probability (0.2) of it moving to fine heavy polishing.

Whenever the process is executed, the worker and nature make a series of choices, which result in a path executed on the BPMN with a total impact vector for that specific instance. Let’s now assume that, for economic reasons, the process must stay within a certain bound. Therefore, our interest is always to stay below that bound. However, we have to consider that the path also depends on the natures within the process, of which we do not know the choice a priori, but we only have the probability of going one way or the other. Consequently, we can formulate a strategy, defined as a series of choices taken while considering the nature and a maximum expected impact, to manage to reach the end of the process with a certain impact vector.

\textit{Strategy example:}
after cutting the metal piece, we have two tasks after the parallel split node, so we do the bending and milling in parallel. Then, after milling we have two options to choose from, here we choose fine deposition. After bending, we have two options to choose from: we choose light polishing with the probability of 0.8. Then, we have two final tasks to choose from: we select LPLS painting.
Finally, we have the maximum expected impact of $\scriptsize \begin{bmatrix}
    115 \\
    11 \end{bmatrix} \times 0.2 + 
    \begin{bmatrix}
    135\\
    8
    \end{bmatrix} \times 0.8 =  
    \begin{bmatrix}
    131\\
    8.6
    \end{bmatrix} $
.

A strategy is defined as winning only if the expected impact vector is below the bound. Therefore, the goal is to find a winning strategy. Consider, for example, that you want to keep the \CPI\ visible in Figure \ref{fig:motivatingexample} under the limit of $ei=$ 
$\scriptsize \begin{bmatrix}
155 \\
7.5 \end{bmatrix}$.
In this case, the strategy shown is not a winning strategy.
In fact, it presents a maximum expected impact greater than the bound $ei$. 
Below we propose an example of a winning strategy.

\textit{Wining strategy example:}
after cutting we perform milling in parallel with bending. we have two options that come after milling; we choose fine deposition. We have two options to choose from after bending; we choose light polishing with a probability of 0.8. Then, we have two final tasks to choose from and select HPHS painting this time.
Finally, we have $\scriptsize \begin{bmatrix}
    135 \\
    9 \end{bmatrix} \times 0.2 + 
    \begin{bmatrix}
    155\\
    6
    \end{bmatrix} \times 0.8 =  
    \begin{bmatrix}
    151\\
    6.6
    \end{bmatrix} \leq ei$
, so this strategy successfully keeps the overall impact below the expected impact.

\subsection{Problem Formulation}
\label{subsec:problem}

\begin{figure}[t]
    \centering
    \begin{tikzpicture}
    \node[label={[]270:(a)}](A){\includegraphics[scale=0.45]{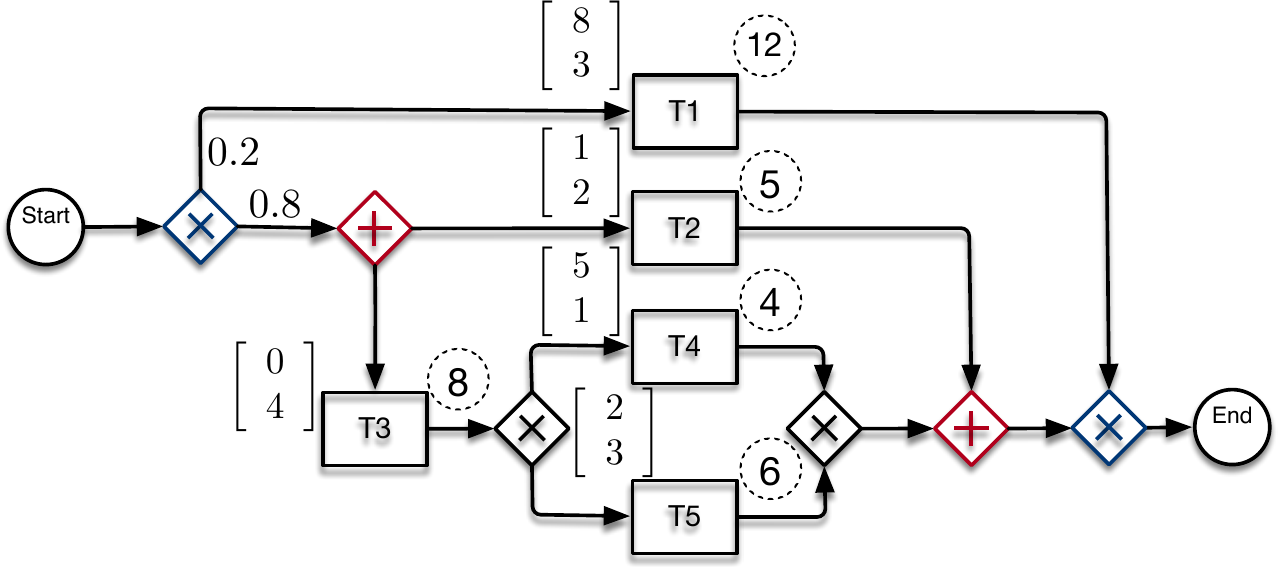}};
    \node[anchor=north, label={[]270:(b)}, xshift=0cm](B) at (A.south){\hstretch{1}{\includegraphics[scale=0.45]{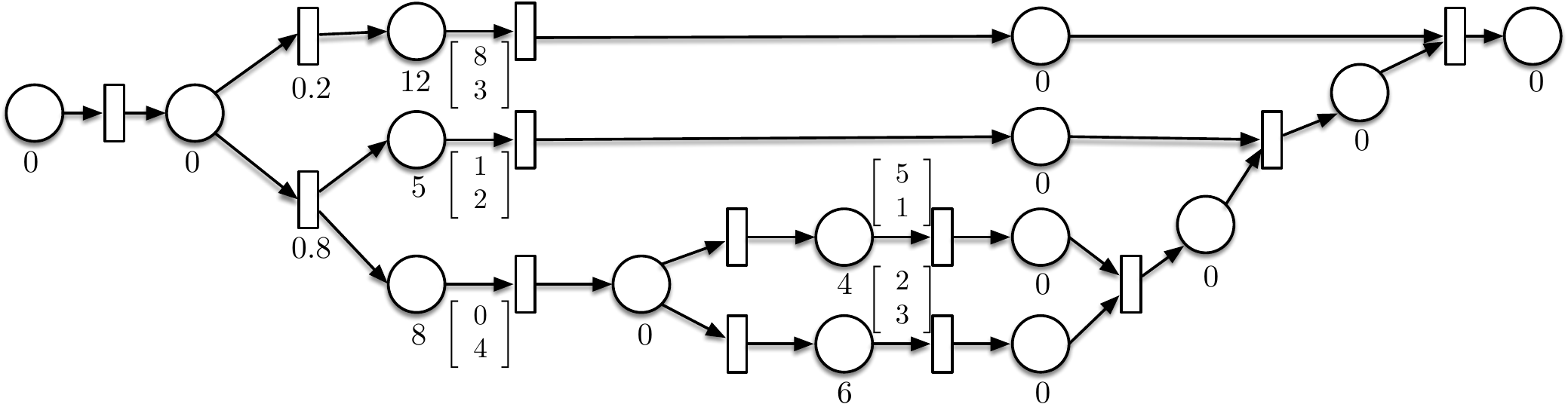}}};
    
    \end{tikzpicture}
    
    \caption{A \CPI\ utilizing all the components considered in this work (a) and its $\spin$ translation (b).}
    \label{fig:bpmn}
\end{figure}

In this section, we formally state the 
\CPI\ semantics.
First, we define the concept of Structured Single-Entry Single-Exit (SESE) BPMN, Figure~\ref{fig:bpmn}, as follows.
\begin{definition}\label{def:structured}
A \emph{structured  single-entry-single-exit diagram}, from now on
simply a SESE diagram, is a
directed graph $\cD = (V, E, E_\top, \cT)$ where 
$(V, E)$ is a directed graph, $E_\top \subseteq E$, $\cT: V \rightarrow \{ event, task, join, \allowbreak split \}$
such that:
\begin{compactenum}

\item for each $v\in V$ if $\cT(v) = event$ then 
there exists at most one edge departing from $v$,
there exists at most one edge entering $v$, and at least one edge
departing from $v$ or entering $v$, i.e., 
$|\{(v,v') \in E\}| \leq 1$,
$|\{(v',v) \in  E\}| \leq 1$,
and $|\{(v',v) \in  E\} \cup \{(v,v') \in  E\}| > 0$;

\item there exists exactly two distinct 
nodes $\hv,\cv$ in $V$ such that $\hv$ has not incoming edges
and $\cv$ has not outgoing edges, i.e., 
$\{(v,\hv) \in E\}=\{(\cv,v) \in E\}=\emptyset$;

\item for each $v\in V$ if $\cT(v) = task$
there exists exactly one edge departing from $v$
and one edge entering $v$, i.e.,  $|\{(v,v') \in E\}|=|\{(v',v) \in E\}| = 1$;

\item for each $v\in V$ if $\cT(v) = split$
there exists exactly two edges departing from $v$
and one edge entering $v$, i.e., 
$|\{(v,v') \in E\}|= 2$ and $|\{(v',v) \in E\}| = 1$;

\item $E_\top \subseteq \{ (v,v'): \cT(v) = split \}$ and for each $v\in V$ if $\cT(v) = split$ we have $|\{v': (v,v') \in E_\top \}|  = 1$;

\item for each $v\in V$ if $\cT(v) = join$
there exists exactly one edge departing from $v$
and two edges entering $v$, i.e., 
$|\{(v,v') \in E\}|= 1$ and $|\{(v',v) \in E\}| = 2$;

\end{compactenum}

\end{definition}

Every non-SESE BPMN diagram can be translated into a SESE diagram as demonstrated in~\cite{DBLP:conf/bpmn/DumasGP10}.

In particular, in the rest of this work, we will restrict ourselves to acyclic SESE diagrams. We will discuss this limitation and how it can be overcome in 
Section~\ref{subsec:loops}.


We define \CPI\ processes as follows.

\newcommand{\pe}{Pcpi}

\begin{definition}\label{def:cpi}
A \CPI\ is a tuple $\pe = (\cD, \cP, \cI, \delta)$
where $\cD = (V, E, \cT)$ is a SESE diagram,
and $\cP: split(V) \rightarrow \bbR_{[0,1]}$ is a \emph{partial function}, $\cI: task(V)\rightarrow (\bbR_{\geq_0})^k$
with $k \in \bbN$, and $\delta: task(V)\rightarrow \bbN^+$.
\end{definition}

Let us notice that since  $\cP$ is a partial function, it 
suffices to encode the natural split gateways in the diagrams, i.e., the one with associated probabilities.
Then  we may define $V_{nature}$ as the set $V_{nature}= Dom(\cP)$ and, on the other hand, for the choice of the system 
$V_{choice}$ as $V_{choice} = split(V) \setminus V_{nature}$.

Let us now extend the semantics of classical Petri nets~\cite{peterson1977petri} in order to capture the semantics of \CPI\ process.

\begin{definition}\label{ref:spinet}
    A Simultaneous Probabilistic Impactful
    Network ($\spin$) is a tuple $N = (PT = P \cup T, 
    T_p, \Delta,\allowbreak I, Pr, D)$ where $P$ and 
    $T$ are finite disjoint set of places
    and transition, respectively, $T_p \subseteq T$, 
    $\Delta \subseteq (P \times T) \cup (T \times P)$, 
    $I: T \rightarrow \bbN^k$, $D: P \rightarrow\bbN$,
    and $Pr: T_p \rightarrow [0,1]$.
\end{definition}

Given a $\spin$  $N = (PT = P \cup T, 
    T_p, \Delta, I, Pr, D)$
for each $pt \in PT$ let $\incoming(pt)= \{pt' \in PT: (pt', pt) \in \Delta\}$ and let   $\outgoing(pt)= \{pt' \in PT: (pt, pt') 
\in \Delta\}$.  
Here we focus on a specific restriction of $\spin$ called 
\emph{structured acyclic $\spin$}.

\begin{definition}\label{def:structuredspin}
We say that a $\spin$ 
$N = (PT= P \cup T, T_p, \Delta, I, Pr, D)$ is 
\emph{structured and acyclic} if and only if the directed graph $(PT, \Delta)$ is acyclic, and the following conditions hold:
\begin{compactenum}

    \item for each $pt \in PT$ we have, $|\incoming(pt)|\leq 2$
    $|\outgoing(pt)| \leq 2$ and $|\incoming(pt)| + |\outgoing(pt)|\leq 3$;
    
    \item there exists a unique partition $\cT_p = \{t_1, \overline{t}_1\}, \ldots,\{t_m, \overline{t}_m\}$ 
of $T_p$ such that  $Pr(t_i) = 1 - Pr(\overline{t}_i)$, \\
$|\outgoing(t_i)| =|\outgoing(\overline{t}_i)| = 1$,
and $\incoming(t_i) = \incoming(\overline{t}_i)$;

    \item there exists a unique set cover $PT_1, \ldots PT_m$  
    of $PT$ such that for each pair $PT_i, PT_j$ 
    of the cover $PT_i\cup PT_j$ also belongs to the cover and
    the following conditions hold:
    \begin{compactitem}

        \item  $pt \in PT$ we have that there exists at most two incoming and two outgoing edges, and the cardinality of the incoming and outgoing edges is at most $3$, i.e., 
        $\{(pt', pt)\}$.
        
        \item for each pair $PT_i, PT_j$
        $PT_i \cap PT_j = \emptyset$, or
        $PT_i \subseteq PT_j$, or $PT_j \subseteq PT_i$;
    
        \item for each $PT_i \neq PT$ there exists a unique 
        element $pt_{in(i)} \in PT_i$ (resp., $pt_{out(i)} 
        \in PT_i$) such that  $\{pt_{in(i)}\} =
        \{pt: (pt',pt) \in \Delta, pt' \notin PT_i, pt \in PT_i \}$ (resp., $\{pt_{out(i)}\} =
        \{pt: (pt,pt') \in \Delta, pt' \notin PT_i, pt \in PT_i \}$);

        \item for each $PT_i \neq PT$  all the elements  of $PT_i$ are reachable  from  $pt_{in(i)}$  via $\Delta$
        and all the elements of $PT_i$ can reach $pt_{out(i)}$
        via $\Delta$.

    \end{compactitem}

\end{compactenum}
 
\end{definition}
The class of $\spin$, as captured by Definition~\ref{def:structuredspin}, is the counterpart of acyclic \CPI. The formal translation from \CPI\ to $\spin$\ provided which enriches the work~\cite{dijkman2008semantics}, is not here shown for the sake of brevity. However, an example that includes the main BMPN elements is shown in Figure~\ref{fig:bpmn}.

Let us notice that by the above definition a 
structured acyclic $\spin$, a $\spin$ from now on,
features exactly one place $p_0$ with $\incoming(p_0) = \emptyset$
and a unique place $p_f$ with $\outgoing(p_f) = \emptyset$.
Let us define a 
switch function $sw: T_p \rightarrow T_p$  
such that for every $t \in T_p$ $\{sw(t),t'\} \in \cT_p$.
Basically $sw$ act as a tool that allow us, for every probabilistic transition $t$,
to access the unique other probabilistic $\overline{t}$ transition which shares the same incoming place of $t$.

Let us now formally define how 
computations work for $\spin$s. 
Given a $\spin$ 
$N = (PT= P \cup T, T_p, \Delta, I, Pr, D)$,
a state $q: P \rightarrow\bbN\cup\left\{\epsilon\right\}$ is a function that maps places in temporal units, where $\epsilon$ states that the specific place has not been visited yet, or that it has already been visited. 

\noindent \begin{tabular}{l} Initial state $q_0$ and final state $q_f$\\ for a $\spin$ are defined as follows:
\end{tabular}
\begin{tabular}{ccc}
$q_0(p) =
\begin{cases}   0 & \mbox{if $p = p_0$} \\
                \epsilon & \mbox{otherwise} \\
\end{cases}$ &; &
$q_f(p) =
\begin{cases}   0 & \mbox{if $p = p_f$} \\
                \epsilon & \mbox{otherwise} \\
\end{cases}$
\end{tabular}.

We will say that a transition $t \in T$ is enabled in a state $q$ if and only if, for all $p \in \incoming(t)$, $q(p) \geq D(p)$.
Let us introduce now the concept of saturated state.

\begin{definition} \label{def:saturated}
Given a state $q$ for a spin
$N = (PT= P \cup T, T_p, \Delta, I, Pr, D)$
we say that $q$ is saturated
if and only if there exists at least one transition $t\in T$ which is enabled in $q$
\end{definition}

Since in a not saturated state  $q$ no transition $t \in T$ is enabled the net will be stuck in $q$. Then the intuition behind not saturated states is that the corresponding \CPI\ process is waiting for one or more tasks to terminate before going further. 
For getting out of such not saturated states we introduce a special transition $t_w$, the so called \textit{wait transition} which encode the passing of one time units and it is enabled only in not saturated states. 

Unlike classical Petri Nets, where each transition is fired one at the time here may fire either $t_w$ or a subset of $T$ called maximal non-conflicting enabled transition set.

\begin{definition}\label{def:mnce}
Given a state $q$ for a spin
$N = (PT= P \cup T, T_p, \Delta, I, Pr, D)$ and a subset $\overline{T} \subseteq T$ we say that $\overline{T}$ is a \emph{maximal non-conflicting enabled 
transition set}, $\mnce$ for short, in $q$ if and only if the following conditions hold:
\begin{compactenum}
\item  for each $t \in \overline{T}$ we have that $t$ is enabled in 
$q$
(\emph{enabled});
\item  for each $t, t'\in \overline{T}$ with $t\neq t'$
we have $(\incoming(t)\cup \outgoing(t)) \cap 
(\incoming(t')\cup \outgoing(t')) = \emptyset$ (\emph{non-conflicting});
\item for any $t \in T \setminus \overline{T}$ 
we have that $\overline{T} \cup \{t\}$ violates 
the above two conditions
(\emph{maximal}); 
\end{compactenum}

\end{definition}

\begin{wrapfigure}{r}{0.5\textwidth}
\vspace{-0cm}
    \centering
    \includegraphics[scale = 0.4]{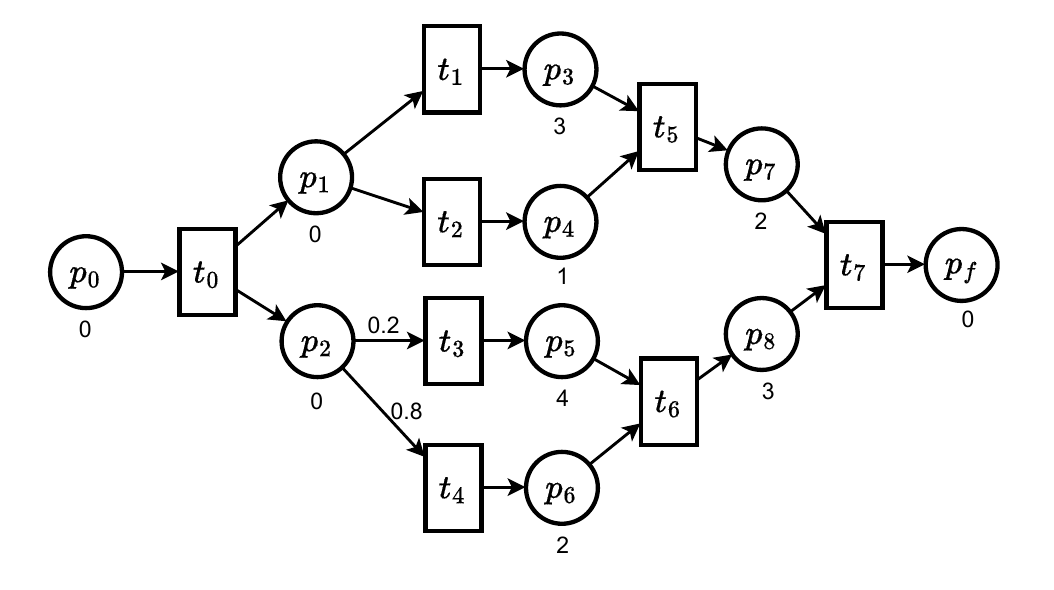}
    \vspace{-0.5cm}
    \caption{A $\spin$ for illustrating $\mnce$ and probabilistic variants.}
    \label{fig:transitiondiagram}
\end{wrapfigure}

\noindent 
Given a set of transitions \(\overline{T} \subseteq T\), let

\[
\outplaces(\overline{T}) = \bigcup_{t \in \overline{T}} \outgoing(t)
\]

and let

\[
\places(\overline{T}) = \bigcup_{t \in \overline{T}} \incoming(t) \cup \outplaces(\overline{T}).
\]

\noindent Now we are ready to define the transition relation between states in a $\spin$. Let $N = (PT= P \cup T, T_p, \Delta, I, Pr, D)$ 
a spin for any pair of states $q$,$q'$ for it we have:\\
\noindent
\begin{tabular}{ccc}
\hspace{-0.3cm}
\begin{tabular}{ccc}
$q \ensuremath{{\mathop{\rightarrow}\limits^{_{t_w}}}} q'$ 
&\hspace{-0.3cm}
\textbf{iff}\hspace{-0.3cm}
&
\begin{tabular}{c}
$q$ is not saturated \\ and \\
$q'(p) =
\begin{cases}   q(p)+1 & \mbox{if $q(p) \in \bbN$} \\ 
                \epsilon & \mbox{otherwise} \\
\end{cases}$

\end{tabular}

\end{tabular}
\! \hspace{-0.5cm}

& 
;
&
\begin{tabular}{ccc}
$q \ensuremath{{\mathop{\rightarrow}\limits^{_{\overline{T}}}}} q'$ 
&\hspace{-0.3cm}
\textbf{iff}\hspace{-0.3cm}
&
\begin{tabular}{c}
$q$ is  saturated, $\overline{T}$ is an \mnce\ in $q$, \\ and \\
$q'(p) =
\begin{cases}   q(p)+1 & 
\begin{array}{c}
\mbox{ if } q(p) \in \bbN  \mbox{ and } \\
p \notin \places(\overline{T})
\end{array}
\\
                0 & \mbox{if $p \in \outplaces(\overline{T})$} \\
                \epsilon & \mbox{otherwise} \\
\end{cases}
$

\end{tabular}

\end{tabular}
\hspace{-0.5cm}.
\end{tabular}

\begin{definition}\label{ref:spincomputation}
    A computation $c = q_0 \ensuremath{{\mathop{\rightarrow}\limits^{_{\overline{T_1}}}}} \ldots \ensuremath{{\mathop{\rightarrow}\limits^{_{\overline{T_n}}}}} q$ in a $\spin$ is a sequence of sets of transitions $\overline{T_i}$ where for each $1\leq i \leq n$ we have that $\overline{T_i}$ is either $t_w$ or an \mnce\ for $q_{i-1}$.
\end{definition}

\noindent A computation $c = q_0 \ensuremath{{\mathop{\rightarrow}\limits^{_{\overline{T_1}}}}} \ldots \ensuremath{{\mathop{\rightarrow}\limits^{_{\overline{T_n}}}}} q$ in a $\spin$ is called a final computation if $q = q_f$.
Stated that $I(t_w) = 0^k$ we can compute $I(c) = \sum\limits_{t \in \bigcup_{i=1}^{n} \overline{T_i}} I(t)$ the impact associated with the computation $c$ and $p(c) = \prod\limits_{t \in \bigcup_{i=1}^n \overline{T_i}\cap T_p} Pr(t)$, the probability associated with the computation $c$. Let ${T}_{\cancel{p}} = T \setminus T_p$, that is, the set of transitions devoid of probabilistic transition, a strategy is defined as follows.

\begin{definition}\label{ref:strategy}
    Let $\mathbb{C}$ be the set of all the computations for a $\spin$, we can define a strategy $S: \mathbb{C} \rightarrow 2^{{T}_{\cancel{p}}} \cup \{t_w\}$, a function that maps computations either into subsets of ${T}_{\cancel{p}}$ or into $t_w$.
\end{definition}

\noindent So, starting from a computation $c$ in which we have reached the last state of the sequence, a strategy $S(c)$ tells us which are the next non-probabilistic transitions that are going to be \textit{fired}.  For all computations $c = q_0 \ensuremath{{\mathop{\rightarrow}\limits^{_{\overline{T_1}}}}} \ldots \ensuremath{{\mathop{\rightarrow}\limits^{_{\overline{T_n}}}}} q$
we implicitly assume that $S(c)$ is $t_w$ if $q$ is not saturated 
and for and does not exists an enabled
transition $t \in T_{\cancel{p}} \setminus S(c)$
such that $t\cup S(c)$ is non-conflicting, i.e., 
$S(c)$ may always be completed into an \mnce\ for $q$. Given a computation $c = q_0 \ensuremath{{\mathop{\rightarrow}\limits^{_{\overline{T_1}}}}} \ldots \ensuremath{{\mathop{\rightarrow}\limits^{_{\overline{T_i}}}}} q_i \ensuremath{{\mathop{\rightarrow}\limits^{_{\overline{T}_{i+1}}}}} \ldots \ensuremath{{\mathop{\rightarrow}\limits^{_{\overline{T_n}}}}} q$, we refer to the first $i$ transitions sets of the sequence with the term sub-computation, written $c_{[0 \ldots i]}$.

\begin{definition}\label{ref:playofS}
    Given a strategy $S$, a \textit{play of S} is a computation $c = q_0 \ensuremath{{\mathop{\rightarrow}\limits^{_{\overline{T_1}}}}} \ldots \ensuremath{{\mathop{\rightarrow}\limits^{_{\overline{T_n}}}}} q$, such that for all sub-computations $c_{[0 \ldots i]}$, $S(c_{[0 \ldots i]}) \in \overline{T}_{i+1}$. 
\end{definition}

\noindent Let $Games(S)$ be the set of all the \emph{final} computations in $\mathbb{C}$ which are also plays of $S$.

\begin{definition}\label{ref:winningStrategy}
    Given a vector bound $\mathbb{EI} \in \bbN ^k$, a strategy $S$ is said to be winning for $\mathbb{EI}$ if and only if $\sum\limits_{ c \in Games(S)} p(c)I(c) \leq \mathbb{EI}$. 
\end{definition}

\noindent Finally, we highlight the problem we aim to resolve throughout this work.

\begin{tcolorbox}
\begin{problem}\label{prob:winningStrategyExistence}
    Given a structured acyclic $\spin$ and an expected vector bound $\mathbb{EI}$ decide whether or not there exists a winning strategy $S$ for $\mathbb{EI}$ in $\spin$.
\end{problem}
\end{tcolorbox}

Given a generic state 
$p$, e.g., $q = \{ p_1 \mapsto 0, p_2 \mapsto 0\}$ (for the sake of brevity, because the other positions are equal to $\epsilon$ are not inserted in $q$)  from the diagram in Figure~\ref{fig:transitiondiagram}.
Then, we are interested in the \mnce\ set of transitions and suppose we are in state $q$. 
In this case, all the \mnce\ are $ \{t_1, t_3\}, \{t_1, t_4\}, \{t_2, t_3\}, \{t_2, t_4\}$.
Now, consider a transition set $\overline{T}$ where $\overline{T} = \{ t_1\}$; it is clear that $\overline{T}$ is not \mnce\ because it is not maximal as it does not consider a transition that originates from $p_2$, e.g. can be extended to $\{ t_1, t_2\}$.
Let's suppose now that we have $\overline{T} = \{ t_1, t_4, t_5\}$. In this case, it is not \mnce\ because it contains $t_5$ that is not enabled. 
Finally, let's suppose that we have $\overline{T} = \{ t_1, t_2, t_4\}$; it is not \mnce\ because it contains $t_1, t_2$ that have the same origin in $p_1$. In fact, $\incoming(t_1) = \incoming(t_2) = p_1$.
We now propose an example to clarify what we mean by $Pvariant(\overline{T})~\ref{def:Pvariant}$. 
Consider the \mnce\ $\{t_1, t_3\}$, the $Pvariant$ is $Pvariant(\{t_1, t_3\}) = Pvariant(\{t_1, t_4\}) = \{ \{t_1, t_3\}, \{t_1, t_4\}\}$. Notice that one variant of $\{t_1, t_3\}$ is $\{t_1, t_3\}$, because one variant of $\overline{T}$ is always $\overline{T}$.

\subsection{Dealing with Loops}\label{subsec:loops}
\begin{figure}
\centering
\hspace{-0.2cm}\includegraphics[scale=0.4]{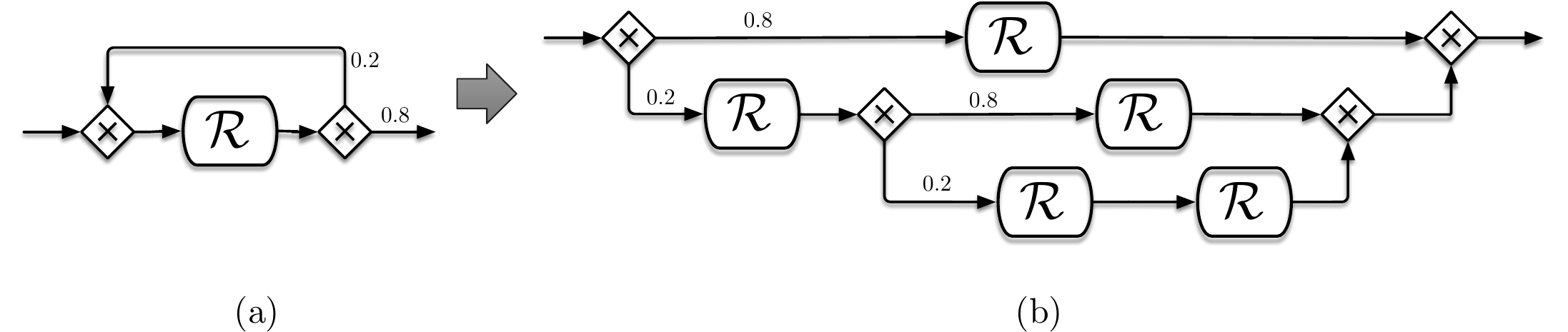}
\caption{\label{fig:loopunravel} 
An example of a $2$-unraveling of a loop region. 
}
\end{figure}

Despite the whole work being based on acyclic SESE diagrams, we are aware that an important component of such diagrams is missing, i.e., loops.
We briefly discuss a simple method for handling loops in our framework, and we are interested in exploring more elegant and theoretical options in future developments.
In our framework, the split node $v$ that induces a loop must be a nature one, i.e., $v \notin V_{nature}$, to represent a more general problem. If $v \in V_{choice}$, we restrict our search to strategies that avoid further loop iterations due to the non-negative nature of impacts.
We highlight a set $ V_{loop} \subseteq V_{nature}$, identifying split nodes encapsulating a loop region. We introduce a function $maxloop: V_{loop}\rightarrow \bbN$ to encode the maximum loop iterations, allowing us to unravel the cyclic structure into an acyclic one.
An example of this unraveling is provided in Figure~\ref{fig:loopunravel}, where $maxloop(v) = 2$ results in a chain of $2$ copies of $v$ nested into each other. Each additional iteration reduces the contribution to the expected impact by an order of magnitude.
This approach is simple to understand and implement and can be parametrized by the user. However, it may result in an exponential increase in size for multiple nested loops, even if $max(Img(maxloop))$ is small. This could affect the feasibility of finding a winning strategy.

We would like to point out that the finite user-parametrized loop unraveling is one of the simplest and most common approaches adopted in the BPMN field \cite{dijkman2008semantics} in order to deal with loops. 
For the time being, our tool (see the end of Section~\ref{subsec:gametranslation} for further) deals with loops by the method described above, which is still good for contexts that do not put to much emphasis
on high numbers of iterations of the loops, for quick experiments, or for comparison with more sophisticated methods to come.

\section{Computational Complexity}\label{sec:complexity}

In this section, we provide a complexity upper bound for
Problem~\ref{prob:winningStrategyExistence}, that is PSPACE, by means of the Algorithm~\ref{alg:pspace}.
The lower bound for the complexity, 
which is within NP-HARD and PSPACE 
(NP-HARD lower bound may be provided by a reduction similar to the one presented in Section~\ref{sec:strategies} for $k$ cost game) is still an open problem. 
First, we have to observe that due to the duration constraints, we may have an exponential number of wait steps if we express such durations in binary. 
However, this may be easily dealt with if we consider the fact that chains of wait transition by their very definition
do not generate possible branching in the computation. Let $Q$ the set of all possible states, we  define a function $sat: Q \rightarrow Q$ as follows: $sat(q) = 
    \begin{cases} 
        q & \mbox{if q is saturated} \\
        sat(q') & \begin{array}{c}\mbox{with $q \mathop{\rightarrow}\limits^{_{t_w}} q'$}
        \mbox{otherwise}
        \end{array}\\
    \end{cases}$.
    
Basically, the $sat$ function take a state $q$ and returns the next saturated state that can be obtained by $q$.
Now we can provide the definition of saturating transition
between two saturated states $q, q'$:\\
\begin{center}
\begin{tabular}{ccc}
$q\mathop{\Rightarrow}\limits^{_{\overline{T}}} q'$
&\hspace{-0.3cm}
\textbf{iff}\hspace{-0.3cm}
&
\begin{tabular}{r}
$q$ is saturated, $\overline{T}$ is an \mnce\ for $q$
and either $q \mathop{\rightarrow}\limits^{_{\overline{T}}} q'$
with $q'$ saturated\\ or there exists $q''$ such that
$q \mathop{\rightarrow}\limits^{_{\overline{T}}} q'$
and $sat(q'') = q'$
\end{tabular}
\end{tabular}
\end{center}

\begin{wrapfigure}{r}{0.5\textwidth}
\vspace{-0.2cm}
    \SetKwProg{Proc}{Procedure}{:}{}
\SetKwFunction{Saturate}{Saturate}
\begin{algorithm}[H] 
\caption{
\label{alg:saturate} Saturate{$(q, N)$} }
\BlankLine
\KwIn{ a state $q$ of a $ \spin\ N = (PT = P \cup T,  T_p, \Delta, I, Pr, D)$ }
\KwOut{$sat(q)$ }
\If{there exists $t \in T$ s.t. $t$ is enabled in $q$}{\Return $q$}
\textbf{let} $\overline{T} \subseteq T$ s.t. for each $t \in \overline{T} $ and for each $ p \in \incoming(t)$ we have 
$q(p) \neq \epsilon$\\
\ForEach{$t \in \oT$}{
$k_t \leftarrow \max\{D(p) - q(p): p \in \incoming(t)  \}$
}
$k \leftarrow n\min\{ k_t: t \in \oT \}$

\textbf{let} $q'$ s.t. $\forall p \in P$
        $q'(p) =
    \begin{cases} 
        \epsilon & \mbox{if $q(p) = \epsilon$} \\
        q(p)+k & \mbox{otherwise} \\
    \end{cases}$\\
    
\Return $q'$\\
\end{algorithm}
\vspace{-0.7cm}
\end{wrapfigure}

It is easy to see that a partial strategy $S$ 
that is defined only on the computations $c$ which end in a saturated
state  $q$ is as good as a complete strategy since there is only one  ``move'' allowed  in a not-saturated state.
The decision algorithm for Problem~\ref{prob:winningStrategyExistence}
makes use of Algorithm~\ref{alg:saturate},
that given a state $q$ computes $sat(q)$
in logarithmic space by means of binary
arithmetic.

\noindent Our decision procedure relies on the following notion of variant for and $\mnce$ .

\begin{definition}\label{def:Pvariant}
    Given an $\mnce$ $\oT$ in $q$ 
    an $\mnce$ $\hat{T}$ in $q$ 
    is a \emph{probabilistic variant}
    of $\oT$ if  the following conditions hold:
    \begin{inparaenum}
        \item $\overline{T} \cap ( T \setminus T_p ) = \hat{T} \cap ( T \setminus T_p )$;
        \item $\forall t \in T_p$ s.t. $t, sw(t) \not \in \overline{T}$ we have $t, sw(t) \not \in \hat{T}$;
        \item $\forall t \in (\hat{T} \cap T_p )$ either $t \in \hat{T}$ or $ sw(t)   \in \hat{T}$.
    \end{inparaenum}    
\end{definition}

Informally speaking, a probabilistic variant for an $\mnce$\ $\oT$ in $q$ is still an $\mnce$\ $\hat{T}$ in $q$ which shares with $\oT$ all the non-probabilistic transitions. Given an $\mnce$\ $\oT$ in $q$, we denote with $Pvariant(\oT, q)$ the set of all and only the probabilistic variants of $\oT$ in $q$. Clearly, we have $\oT \in Pvariant(\oT, q)$.

Algorithm~\ref{alg:pspace} employs a non-deterministic approach to ascertain the existence of a viable strategy for a given instance of Problem~\ref{prob:winningStrategyExistence}. This is achieved by dynamically enumerating all possible plays, thereby maintaining only a single play in memory at any given moment. This method ensures polynomial memory utilization while providing a comprehensive evaluation of potential strategies.

For the sake of brevity, we do not provide the full proof that Algorithm~\ref{alg:pspace} works in polynomial space. However, we informally provide the key arguments of the proof:
\begin{compactitem}
\item Algorithm~\ref{alg:pspace} is non-deterministic because it guesses the correct move (if any) at line 9, where $\oT \cap (T \setminus T_p)$ represents the output of the current strategy;
\item $Saturate$ operates in LOGSPACE and deals with the binary representation of durations for places;
\item Given that $N$ is acyclic, we have that any transition is considered at most for one recursive call to $StrategyExists$. Therefore, the number of nested procedure calls is bounded by $|T|$ since $t_w$ transitions are collapsed via the function $Saturate$;
\item In principle, $|Pvariant(\overline{T}, q)|$ (line 10 of Algorithm~\ref{alg:pspace}) may be of the order of $2^{|T|}$. However, since only one element $\hat{T} \in Pvariant(\overline{T}, q)$ is needed at a time for updating $\overline{rei}$ via the recursive call in the body of the for loop (line 12 of Algorithm~\ref{alg:pspace}), it is possible to set up an enumeration to keep the space polynomial at each step.
\end{compactitem}
Since each play may be represented in polynomial space, we have the following result.

\begin{theorem}\label{def:thm}
Problem~\ref{prob:winningStrategyExistence} is NP-HARD and belongs to the complexity class PSPACE.
\end{theorem}

However, our primary objective is to formulate a strategy rather than merely verifying its existence. Consequently, Section~\ref{sec:strategies} is dedicated to addressing the strategy synthesis problem for \CPI. This section elaborates on the proposed solution, central to the functionality of the effective prototype that we have developed and implemented.

\SetKwProg{Proc}{Procedure}{:}{}
\SetKwFunction{FRecurs}{StrategyExists}%
\begin{algorithm}[t]
\caption{\label{alg:pspace} Recursive Procedure for Solving Problem~\ref{prob:winningStrategyExistence}}
\KwIn{a $\spin$ $N = (P, PT = T\cup T_p, \Delta, I, Pr, D)$
and $\mathbb{EI} \in \bbN^k$}
\KwOut{$ei \in \bbR^k$ with $ei \leq \mathbb{EI}$ if 
there exists a strategy with residual expected impact $ei$, and FAIL otherwise}
\textbf{let} $q_0$ be the initial state of 
$N$\;
\Return \FRecurs{\Saturate$(q_0, N), 0^k, 1, \mathbb{EI}$}
\BlankLine
\Proc{\FRecurs{q, im, cp, rei}}{
    \KwData{A saturated state $q$ of $N$, the value cp of the cumulative probability of the current play,
    $im\in \bbR^k$ the current impact for the play, $rei\in \bbR^k$ the residual expected impact currently available for consumption.}
    \KwResult{$rei \in (\bbR^+)^k$  if 
there exists a strategy from the current state $q$ that that has $rei$ residual w.r.t. $ei$, and FAIL otherwise}
    
    \If{$q$ is final}{
        \If{$rei \not \leq 0^k$}{FAIL}
        \Return $rei - (cp \cdot im) $
    }

    \textbf{let} $\overline{T}$ an $\mnce$ for $q'$
    
    $\overline{rei} \leftarrow rei$
    
    \ForEach{$\hat{T} \in Pvariant(\overline{T}, q)$}{
         
         \textbf{let} $q'$ s.t. $q  \stackrel{\hat{T}}{\rightarrow} \hat{q'}$\\
        $\overline{rei} \leftarrow StrategyExist(\Saturate{q', N}, im + \sum_{t \in \hat{T}} I(t), cp \cdot \prod_{t \in \hat{T} \cap T_p} Pr(t), \overline{rei})$
    
        \If{$rei \not \geq 0^k$}{FAIL}
    }
    \Return $\overline{rei}$
}

\end{algorithm}

\noindent The exact complexity of Problem~\ref{prob:winningStrategyExistence}
is still open, we know that it can be proved to be NP-HARD
by means of a reduction from the Partition problem  
introduced in Section~\ref{sec:strategies} for $k$-cost reachability 
games.

The NP-HARD lower bound may be achieved by building a game devoid of nature nodes in a way that resembles the one-player restriction of the generalized game proposed in \cite{fijalkow2010surprizing}, but here Partition is used instead of SAT as the NP-HARD problem we reduce from. In \cite{fijalkow2010surprizing}, the authors
provide a QSAT reduction for the unrestricted case, thus obtaining a 
PSPACE-HARD lower bound. Such a reduction is not directly applicable 
in our setting since our winning conditions embrace all possible plays, not a single one. In other words, 
in \cite{fijalkow2010surprizing}, a faulty strategy may be detected by witnessing a faulty single play it generates, 
while in our setting, a faulty strategy may be detected only by considering a subset (possibly all) the plays it generates. 
For this reason, at this point, we cannot  conjecture the   
exact lower bound for the complexity of 
Problem~\ref{prob:winningStrategyExistence} without further analysis.


\section{Synthesizing Strategies}\label{sec:strategies}
In this section, 
we will take advantage of the $\spin$ translation which has been fully described in Section~\ref{subsec:problem}.
This tree has the foundational semantics of classical Petri Nets~\cite{peterson1977petri} for BPMN process.
These concepts serve as the mathematical and logical basis for describing a graph-game representation and how the strategy is discovered presented below. 
\subsection{A $k$-cost Reachability Game} \label{subsec:kcostgame}
In this section, we will introduce a graph-game representation 
for dealing with the synthesis of strategies given a BPMN+{CPI} diagram $\cD = (V, E, E_\top, \cT)$ which decides whether there exists a strategy that guarantees that the expected
impact of a diagram is dominated by a given impact vector bound $\mathbb{I}$.

\begin{definition}\label{def:board}
A $k$-cost game board is a tuple $\cB = (P = P_\circ \cup P_\square,p_0,F, \cC,M )$ such that 
$p_0 \in P$, $M \subseteq P \times P$, $F \subseteq P$ with $\{(m, m'): m \in F\}= \emptyset$ (i.e, there aren't outgoing edges from $F$),  $\cC: P \rightarrow \bbR^k$, $(P,M)$ is a directed acyclic graph.
\end{definition}

\begin{definition}\label{def:strategy}
Given $k$-cost game board $\cB = (P = P_\circ \cup P_\square,p_0,F, \cC,M )$
a strategy is a function 
$s:P^* \rightarrow P$ such that:
for every $\rho \in P^*$ we have 
$(\rho[-1],s(\rho)) \in M$.
\end{definition}

\begin{definition}\label{def:generatedplays}
Given a $k$-cost game board $\cB = (P = P_\circ \cup P_\square,p_0,F, \cC,M )$
and a strategy $s$, a successful play $\rho\in P^*$ is generated  by $s$ in $\cB$ if and only 
if:
\begin{inparaenum}[(i)]
    \item $\rho[0] = p_0$;
    \item $\rho[-1] \in F$;
    \item for every $0 < i < |\rho|$
    if $\rho[i - 1] \in P_\circ$
    then $\rho[i]= s(\rho[0:i])$.
\end{inparaenum}
\end{definition}

Let $P^*_s$ be the set of all the
possible plays generated by $s$.

\begin{definition}\label{def:closedplays}
Given $s$ we say that  $P^*_s$ is 
closed if for each $\rho \in P^*_s$
and for each $0 \leq i < |\rho| - 1$
such that   $\rho[i]\in P_\square$
then for each $(\rho[i], p) \in M$
we have that there exists 
$\rho' \in P^*_s$ with $\rho'[0:i] = \rho[0:i]$ and $\rho'[i+1]= p$.
\end{definition}

Given a $P^*_s$ we let
$final(P^*_s)$ the set 
$final(P^*_s)=\{\rho[-1]: \rho \in P^*_s \}$.

\begin{problem}\label{prob:kcostgame}
    Given a  $k$-cost game board
    $\cB = (P = P_\circ \cup P_\square, p_0,  F, \cC, M)$ and a cost $c \in \bbR^k$ determine whether or not there exists a strategy $s$ for 
    which $P^*_s$ is closed and 
    $\sum_{p \in final(P^*_s)} \cC(p) 
    \leq c$.
\end{problem}

A strategy $s$ is positional if and only 
if for every $\rho, \rho' \in P^*$ we 
have that $\rho[-1]= \rho'[-1]$ implies 
$s(\rho)=s(\rho')$.
For the purpose of our game, w.l.o.g. a positional 
strategy may be redefined as $s: P_\circ 
\rightarrow P$.

\begin{problem}\label{prob:kcostgamepos}
    Given a  $k$-cost game board
    $\cB = (P = P_\circ \cup P_\square, p_0,  F, \cC, M)$ and a cost $c \in \bbR^k$ determine whether or not there exists a \textbf{positional} strategy $s$ for 
    which $P^*_s$ is closed and 
    $\sum_{p \in final(P^*_s)} \cC(p) 
    \leq c$.
\end{problem}

\begin{wrapfigure}{r}{0.6\textwidth}
    \centering
    \includegraphics[scale=0.38]{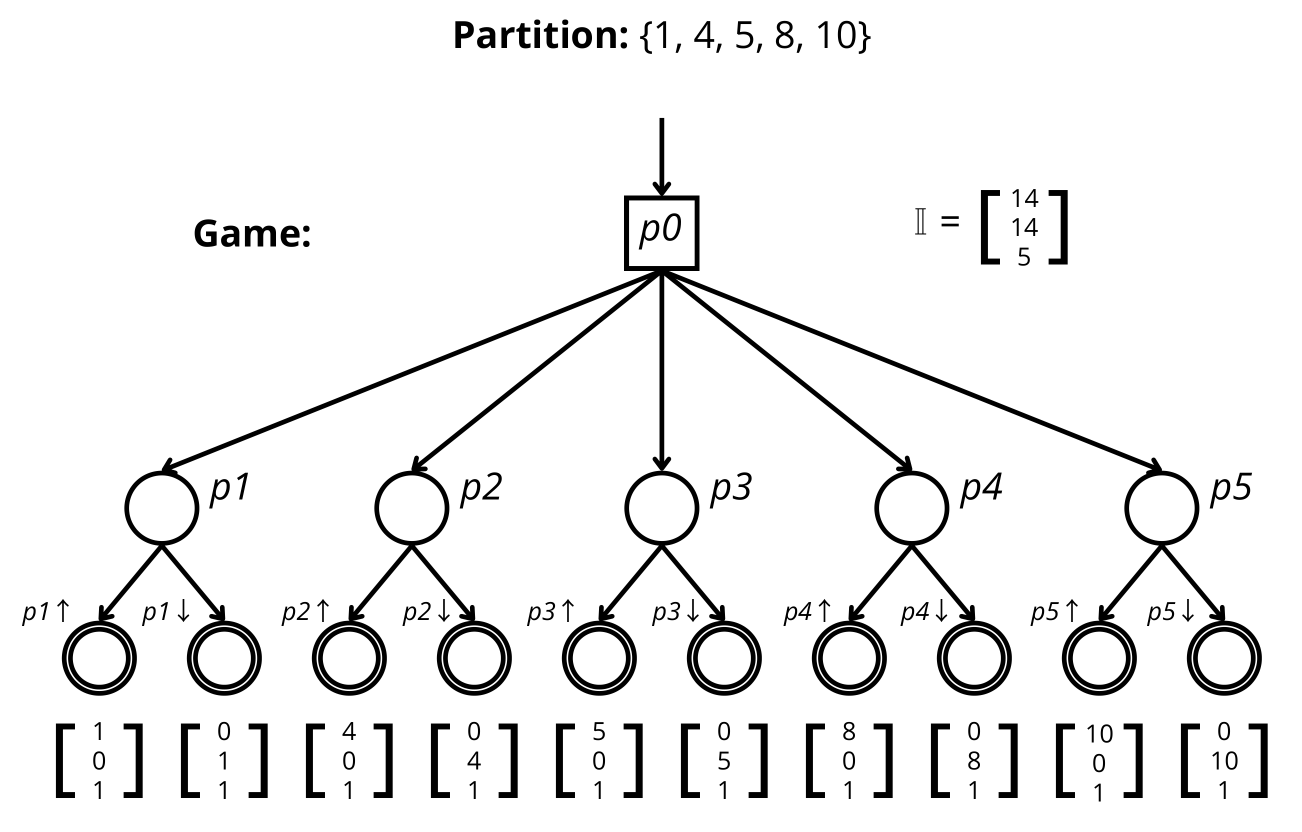}
    \caption{Reduction from Partition to k-cost game Problem.}
    \label{fig:partitionTogame}
    \vspace{0.5cm}
\end{wrapfigure}

\begin{theorem}\label{thm:positional}
For every $k$-cost game board and each 
cost vector $c \in \bbR^k$ 
we have that
$(\cB, c)$ is a positive instance of 
Problem~\ref{prob:kcostgame}
if and only if $(\cB, c)$
is a positive instance of 
Problem~\ref{prob:kcostgamepos}
\end{theorem}

It is easy to prove that Problem~\ref{prob:kcostgamepos}
belongs to the complexity class NP, by simply provide a succinct 
certificate, that is, given an instance  
$(\cB = (P = P_\circ \cup P_\square,M, p_0,  F, \cC), c)$  of 
Problem~\ref{prob:kcostgamepos}
guess a subset $M' \subseteq M$ such that 
$ \{(p, p') \in M: p \in P_\square \} \subseteq M'$
and for each $p \in P_\circ$
either $\{(p, p')\in M\} = \emptyset$ or
there exists a unique edge $(p,p') \in M'$.
Then, let  $F'$ be the subset of $F$ reachable from $p_0$ in
the $M'$-induced sub-graph
$(P_\circ \cup P_\square,M')$ we have that $M'$ is a solution 
if and only if $\sum\limits_{p \in F'} \cC(p) \leq c$.
The NP-HARD lower bound for Problem~\ref{prob:kcostgamepos}, and thus for 
Problem~\ref{prob:kcostgame}, is proved by a reduction from the following NP-HARD problem.

\begin{problem}\label{prob:partition} \emph{(Distinct Partition)}
Given a set of natural numbers $S = \{n_1, \ldots,$ $ n_m \}$ decide whether or not there 
exists a partition $(S_1, S_2)$ of $S$ such that $\sum\limits_{n \in S_1} n= \sum\limits_{n \in S_2} n$.
\end{problem}

\noindent As formulated by Korf in \cite{KORF1998181}, Problem \ref{prob:partition} is actually NP-complete. We recall this in Theorem \ref{thm:partition}.

\begin{theorem}\label{thm:partition}
Distinct Partition (Problem~\ref{prob:partition}) is NP-Complete \cite{KORF1998181}.
\end{theorem}

\noindent There exists  a simple LOG-SPACE reduction from Distinct Partition to Problem \ref{prob:kcostgamepos}, and thus to Problem \ref{prob:kcostgame},
for $k \geq 3$. The reduction is very simple, it suffices to transform the distinct partition problem $S = \{n_1, \ldots,$ $ n_m \}$ into an instance of  
Problem~\ref{prob:kcostgame}
$ (\cB_S = (P = P_\circ \cup P_\square,M, p_0,  F, \cC), c_S)$ as follows:
\begin{compactenum}
    \item $P_\circ= \{ p^i, p^i_{\uparrow}, p^i_{\downarrow}: 1 \leq i \leq m \}$,
    \item $P_\square = \{p_0\}$, 
    \item $M=\{(p_0,p^i): 1 \leq i \leq m\} \cup 
    \{(p^i,p^i_\uparrow), (p^i,p^i_\downarrow): 1 \leq i \leq m\} $,
    \item  $F = \{ p^i_\downarrow, p^i_\uparrow: 1 \leq i \leq m\}$,
    \item  $\cC(p^i_\uparrow) = [n_i, 0, 1]$ and  $\cC(p^i_\downarrow) = [0, n_i, 1]$
    for each $1\leq i \leq m$,
    \item $c_S= \left[\frac{\sum\limits^m_{1} n_i}{2}, \frac{\sum\limits^m_{1} n_i}{2} ,m\right]$.
\end{compactenum}

An example of the proposed reduction is given in Figure~\ref{fig:partitionTogame}. 
It is easy to prove that  $(\cB_S, c_S)$ is a positive instance of 
Problem \ref{prob:kcostgame} if and only if $S$ is a positive instance of 
the distinct partition problem.

\begin{theorem}\label{def:npcompleteness}
Problem~\ref{prob:kcostgamepos} and Problem~\ref{prob:kcostgame} 
for $k \geq 3$ are NP-Complete problems.
\end{theorem}
\subsection{From \CPI\ to $k$-cost Reachability Game}
\label{subsec:gametranslation}

\begin{figure}[t]
\hspace{-0.9cm}\begin{tikzpicture}

\node[label={[]270:\scalebox{0.7}{(a) Single \mnce, $\oT_\circ \cup \varnothing$}}](A) at (0,1){\includegraphics[scale=0.5]{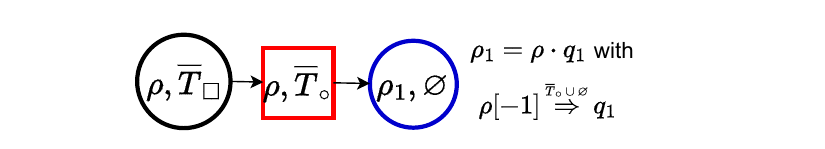}};

\node[label={[]270:\scalebox{0.7}{(b) \begin{tabular}{c}Two \mnce\ one variant of the\\ other,  $\oT_\circ \cup \oT_\square^1 $ and $\oT_\circ \cup \oT_\square^2 $
\end{tabular}
}}](B) 
at (5,0)
{\includegraphics[scale=0.5]{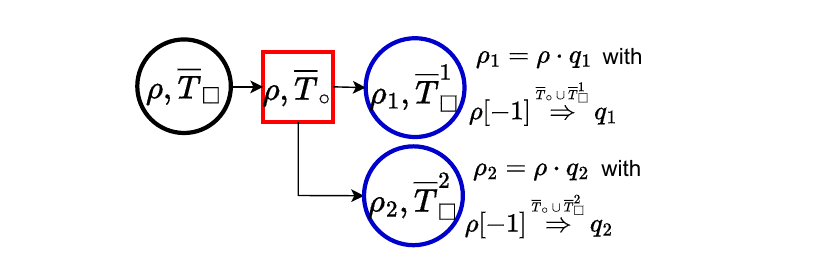}};

\node[label={[]270:\scalebox{0.7}{(c)\begin{tabular}{c} Two \mnce\ representing a choice\\ split with no probabilistic\\ transition, $\oT_\circ^1 \cup \varnothing $ and $\oT_\circ^2 \cup \varnothing$\end{tabular}}}](C)
at (0,-1.5)
{\includegraphics[scale=0.5]{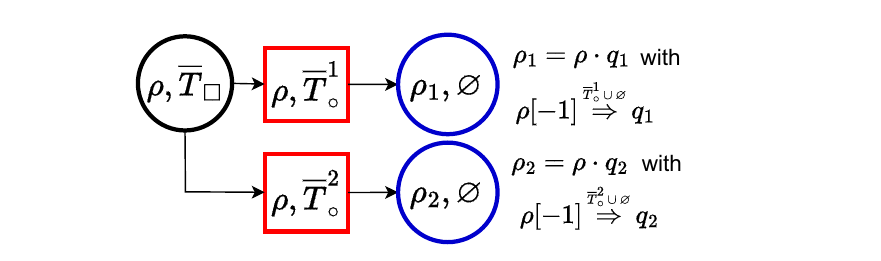}};

\node[label={[label distance=-0.4cm]270:\scalebox{0.7}{(d)\begin{tabular}{c} A choice split happening\\ together with a \\probabilistic split.\end{tabular}}}](D)
at (10,-0.5)
{\includegraphics[scale=0.5]{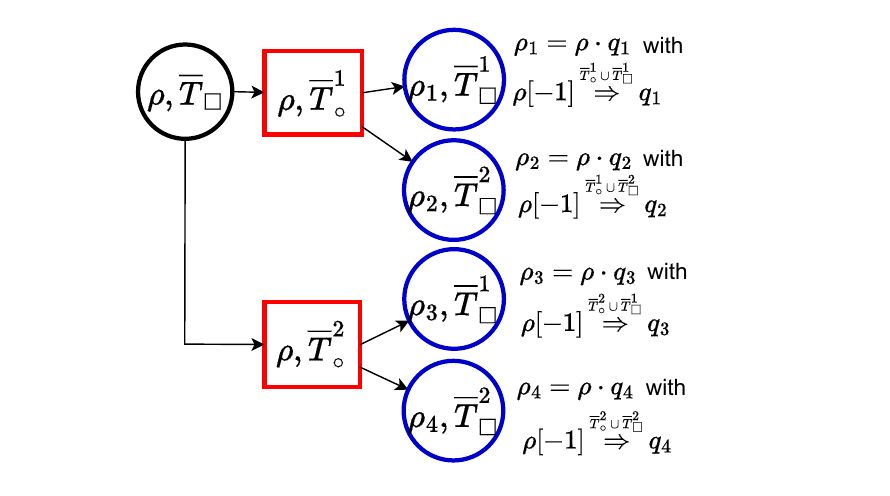}};

\end{tikzpicture}

    \caption{Different scenarios involving at most one choice and at least one probabilistic split.}
    \label{fig:gamesplit1}
\end{figure}

We conclude this section by providing the direct translation from an instance \((N,\mathbb{EI})\) of Problem~\ref{prob:winningStrategyExistence} into a \(k\)-cost game \((\cB,\mathbb{EI})\), which admits a solution if and only if the problem \((N,\mathbb{EI})\) admits a solution. Moreover, if \((\cB,\mathbb{EI})\) admits a solution, i.e., it is a positive instance of Problem~\ref{prob:kcostgame}, such a solution will effectively represent a strategy for the original problem.

Before providing this translation, we introduce a couple of useful definitions. Given an $\mnce$ $\overline{T}$ for a state \(q\), we define two sets: $\overline{T}_\square = \overline{T} \cap T_{p}$ and $\overline{T}_\circ = \overline{T} \setminus T_\square$. Additionally, for any \(\overline{T}_\square \subseteq T_p\), let \(Pr(\overline{T}_\square) = \prod_{t \in \overline{T}_\square} Pr(t)\); clearly, \(Pr(\emptyset) = 1\). For any \(\overline{T}_* \subseteq T\), let \(\cI(\overline{T}_*) = \sum_{t \in \overline{T}_*} \cI(t)\); clearly, \(\cI(\emptyset) = 0\). Finally, let \(Q\) be the set of all possible saturated states on \(N\) and \(C^+_Q\) be the set of all possible non-empty combinations of elements in \(Q\). Given a combination \(\rho \in C^+_Q\), we denote its last element as \(\rho[-1]\).

\noindent Given an instance $(N, \mathbb{EI})$ of 
we define a $k$-cost game board
$\cB_N = (S = S_\circ \cup S_\square,s_0,F, \cC,M )$ 
as follows:\\
\begin{tabular}{cc}\hspace{-0.3cm}
$S_\circ = \{ (\rho, \overline{T}_{\square}) \in C^+_Q \times 2^{T_p}:  \rho[-1] \  is \ saturated \}$,
& \hspace{-0.3cm}
$S_\square = \left\{\hspace{-0.1cm} (\rho, \overline{T}_\circ): \hspace{-0.1cm}
\begin{array}{c}\rho \in C^+_Q, \mbox{ there exists } 
\overline{T}_\square  \subseteq 
T_p  s.t.\\ \ \overline{T}_\circ \cup \overline{T}_\square \ is\ an\ \mnce \ for\  \rho[-1]  \end{array} \hspace{-0.2cm}\right\}$,
\\[0.7cm]
$s_0 = (sat(q_0), \emptyset)$,
&
$F = \{(\rho, \overline{T}_{\square})  \in C^+_Q \times 2^{T_p}: \rho[-1] = q_f\}$, 
\end{tabular}

and $M = \{((\rho, \overline{T}_\square), (\rho, \overline{T}_\circ)): (\rho, \overline{T}_\square) \in P_\circ,
(\rho,P_\square) \in P_\square \} \cup 
\{ ((\rho, \overline{T}_\circ), (\rho q,\overline{T}_\square )):
 \rho[-1]
\mathop{\Rightarrow}\limits^{_{\overline{T}_\circ \cup \overline{T}_\square}} q 
\} $.\\

Graphical examples of how the relation $M$ is build 
in the case when  the \mnce\ $\oT= \oT_{\circ}\cup \oT_{\square}$ satisfies $|\oT_{\circ}| \leq 1$ and $ |\oT_{\square}| \leq 1$ are provided in Figure~\ref{fig:gamesplit1}.

\noindent Lastly, for the cost function, let $M^*$ denote the reflexive and transitive closure of $M$. For any $s \in F$, the cost function $\cC(s)$ is defined as:
\[
\cC(s) = \left(\prod\limits_{(\rho, \overline{T}_\square) \in S_\circ: (s_0, (\rho, \overline{T}_\square)), ((\rho, \overline{T}_\square), s) \in M^*} Pr\left(
\overline{T}_\square 
\right)\right) \cdot \left(\sum\limits_{(\rho, \overline{T}_*) \in S_\circ \cup S_\square: (s_0, (\rho, \overline{T}_*)), ((\rho, \overline{T}_*), s) \in M^*}  \cI\left(\overline{T}_*\right)\right)
\]

\newcommand{\impact}[1]{
\left[\!\begin{array}{l}
#1
\end{array}\!\right]
}

The formula described assigns to each final state $s \in F$ the contribution to the expected impact generated by paths terminating at $s$. Now, as a final measure, we resolve the $k$-cost game by selecting
\footnote{This is implemented by evaluating all possible subsets $F' \subseteq F$ such that $\sum\limits_{s \in F'} \cC(p) \leq \mathbb{EI}$ and for each $s' \in F \setminus F'$, $\sum\limits_{s \in F' \cup \{s'\}} \cC(s) > \mathbb{EI}$. We consider only the maximal admissible subsets of $F$, as they can ``attract'' the initial state if and only if at least one of their subsets does.}
a subset $F' \subseteq F$ such that the total expected impact satisfies: $\sum\limits_{s \in F'} \cC(s) \leq \mathbb{EI}$

We employ the standard attractor procedure as described in \cite{thomas1995synthesis}, initiating with $Attr^0 = F'$ in $\cB$. A positive outcome, along with the strategy formulated by the attractor procedure, is confirmed if there exists $k \in \mathbb{N}$ such that $s_0 \in Attr^k$. While the attractor procedure itself runs in polynomial time, approximately $\mathcal{O}(nm)$ for a graph with $n$ nodes and $m$ edges, the non-deterministic selection of a candidate $Attr^0$ from the set of final states remains computationally intensive, since the number of final states may be exponential in the size of $\spin$ thus
the above procedure for synthetizing 
a strategy operates in NEXPTIME.

\begin{figure}[ht]
    
    \begin{subfigure}[b]{0.3\textwidth}
        \centering
        \includegraphics[scale=0.2]{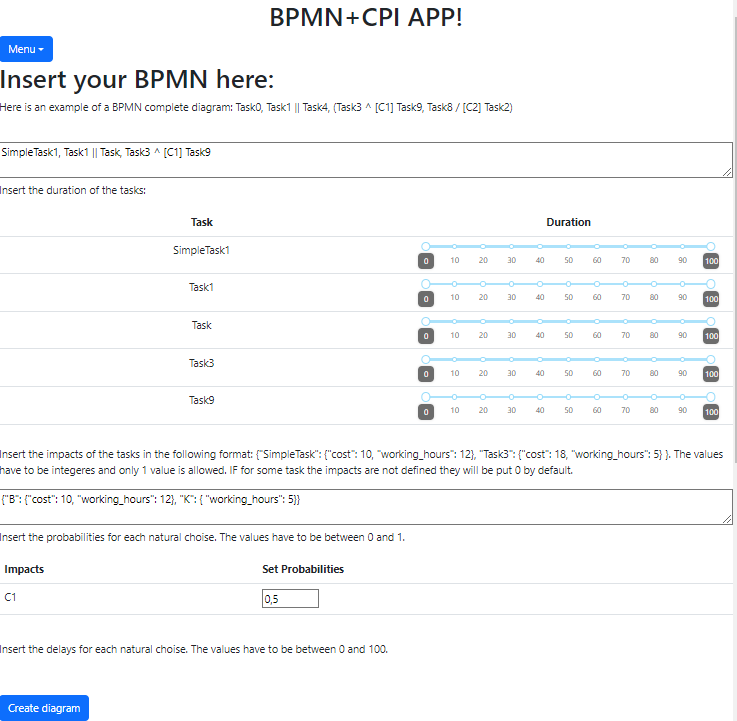}
        \caption{}
        \label{fig:defBPMN}
    \end{subfigure} 
    \hfill
    \begin{subfigure}{0.3\textwidth}        
    \includegraphics[scale=0.3]{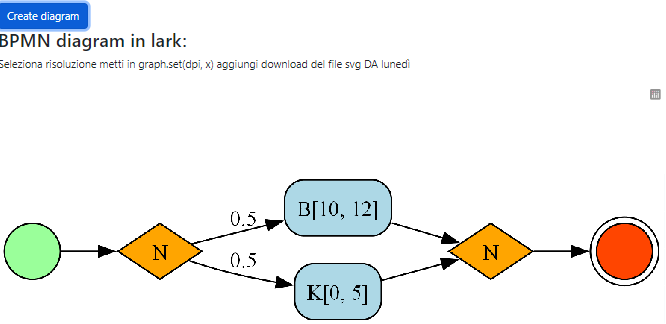}
        \caption{}
        \label{fig:printLarkBPMN}
    \end{subfigure}
    \hfill
    \begin{subfigure}{0.3\textwidth}        
        \includegraphics[scale=0.3]{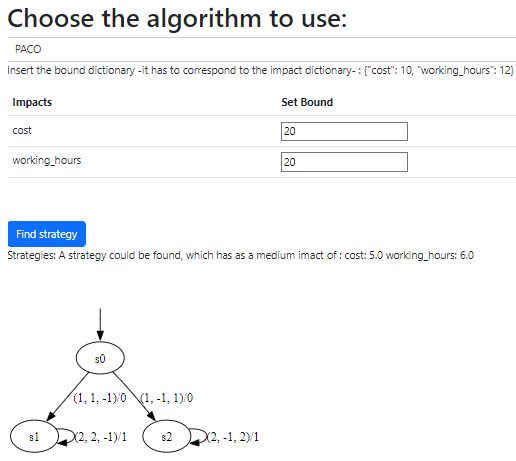}
        \caption{}
        \label{fig:StrategyAppFounded}
    \end{subfigure}
    \caption{Example of using our Dash App: defining the BPMN in our Dah App~\ref{fig:defBPMN}, print the BPMN using Lark~\ref{fig:printLarkBPMN} and example of founded strategy using \emph{PACO}~\ref{fig:StrategyAppFounded}}
    \label{fig:translations}
\end{figure}

\paragraph{Implementation} 
The algorithm described in this section, known as \emph{PACO}, has been developed and is accessible at \url{https://github.com/ansimonetti/PACO}. \emph{PACO} is designed as a Dash App~\cite{dash}. The process is written in Lark syntax~\cite{lark}, with all choices, probabilities, and impacts clearly defined,  as visible in Figure~\ref{fig:defBPMN} and printed using Graphviz~\cite{graphviz} and PyDot~\cite{pydot}, as shown in Figure~\ref{fig:printLarkBPMN}. A specific section is dedicated to defining the expected impacts vector. Subsequently, the AALpy automata~\cite{aalpy} is employed to provide a strategy, as previously described, if one exists. If one is found, the algorithm returns it together with the associated impact factors.
Moreover, it prints the tree associated with the strategy, indicating which tasks have to be done to complete the process within the bound vector as shown in Figure~\ref{fig:StrategyAppFounded}. 
\section{Conclusion}\label{sec:conclusion}

In this study, we developed a BPMN extension, denoted as \CPI, designed to handle execution in the presence of impacts, probabilistic splits, and choices. The semantics for this extension were formulated using an enriched version of Petri Nets, namely, $\spin$. The primary objective of this work was to create a system capable of informing users about the existence of a strategy for a given process and user-defined thresholds. This involves determining whether there is a controller capable of executing each step of the process while ensuring that the expected value of each resource across repeated process instances remains within the predefined thresholds.

First, we proved that the associated decision problem, i.e., determining if such a controller exists, belongs to the complexity class PSPACE. Then, we provided an effective method for building the controller by modifying classical reachability games over graphs. Based on these theoretical results, we implemented a tool capable of determining the existence of a strategy given a \CPI\ process and a given threshold $\mathbb{EI}$. This tool is currently under development, but a working prototype is available online for the benefit of the community.

For future work, we envision two promising extensions. The first, theoretical, aims to deal with loops in the workflow in a non-approximated fashion and to propose alternative algorithms for solving the problem, potentially closing the complexity gap, which currently stands between PSPACE and NP. The second, more practical extension, focuses on better representing the obtained strategy by integrating it into the choice gateway of the \CPI, for instance, representing decisions with a set of inequalities involving intervals of values for the impact components observed in specific choice nodes.

\paragraph{Acknowledgments}
    This work has been carried out while Emanuele Chini was enrolled in the Italian National Doctorate on Artificial Intelligence run by Sapienza University of Rome in collaboration with the University of Verona.
\bibliographystyle{eptcs}
\bibliography{biblio}

\end{document}